\documentstyle[epsfig]{aipproc}
\begin{document}
\input psfig.tex
\title{Optical Observations of Core-Collapse Supernovae}

\author{Alexei V. Filippenko}
\address{Department of Astronomy, University of California,
Berkeley, CA 94720-3411}
\maketitle

\begin{abstract} 
I present an overview of optical observations (mostly spectra) of Type II, Ib,
and Ic supernovae (SNe). SNe~II are defined by the presence of hydrogen, and
exhibit a very wide variety of properties. SNe~II-L tend to show evidence of
late-time interaction with circumstellar material.  SNe~IIn are distinguished
by relatively narrow emission lines with little or no P-Cygni absorption
component and (quite often) slowly declining light curves; they probably have
unusually dense circumstellar gas with which the ejecta interact. Some SNe~IIn,
however, might not be genuine SNe, but rather are ``impostors" ---
specifically, super-outbursts of luminous blue variables. SNe~Ib do not exhibit
the deep 6150~\AA\ absorption characteristic of ``classical" SNe~Ia; instead,
their early-time spectra have He~I absorption lines. SNe~Ic appear similar to
SNe~Ib, but lack the helium lines as well. Spectra of SNe~IIb initially exhibit
hydrogen, yet gradually evolve to resemble those of SNe~Ib; their progenitors
seem to contain only a low-mass skin of hydrogen. Spectropolarimetry thus far
indicates large asymmetries in the ejecta of SNe~IIn, but much smaller ones in
SNe~II-P. As one peers deeper into the ejecta of core-collapse SNe, the
asymmetry (indicated by the amount of polarization) seems to increase.  There
is intriguing, but inconclusive, evidence that some peculiar SNe~IIn might be
associated with gamma-ray bursts. The rates of different kinds of SNe as a
function of Hubble type are still relatively poorly known, although there are
good prospects for future improvement.
\end{abstract}

\section*{ INTRODUCTION}

   Supernovae (SNe) occur in several spectroscopically distinct varieties; see
reference \cite{avf97}, for example.  Type I SNe are defined by the absence of
obvious hydrogen in their optical spectra, except for possible contamination
from superposed H~II regions.  SNe~II all prominently exhibit hydrogen in their
spectra, yet the strength and profile of the H$\alpha$ line vary widely among
these objects.

   The early-time ($t \approx 1$ week past maximum brightness) spectra of SNe
are illustrated in Figure 1. [Unless otherwise noted, the optical spectra
illustrated here were obtained by my group, primarily with the 3-m Shane
reflector at Lick Observatory. When referring to phase of evolution, the
variables $t$ and $\tau$ denote time since {\it maximum brightness} (usually in
the $B$ passband) and time since {\it explosion}, respectively.]  The lines are
broad due to the high velocities of the ejecta, and most of them have P-Cygni
profiles formed by resonant scattering above the photosphere. SNe~Ia are
characterized by a deep absorption trough around 6150~\AA\ produced by
blueshifted Si~II $\lambda$6355. Members of the Ib and Ic subclasses do not
show this line. The presence of moderately strong optical He~I lines,
especially He~I $\lambda$5876, distinguishes SNe~Ib from SNe~Ic.

\bigskip

\hbox{
\hskip +.7truein
\psfig{figure=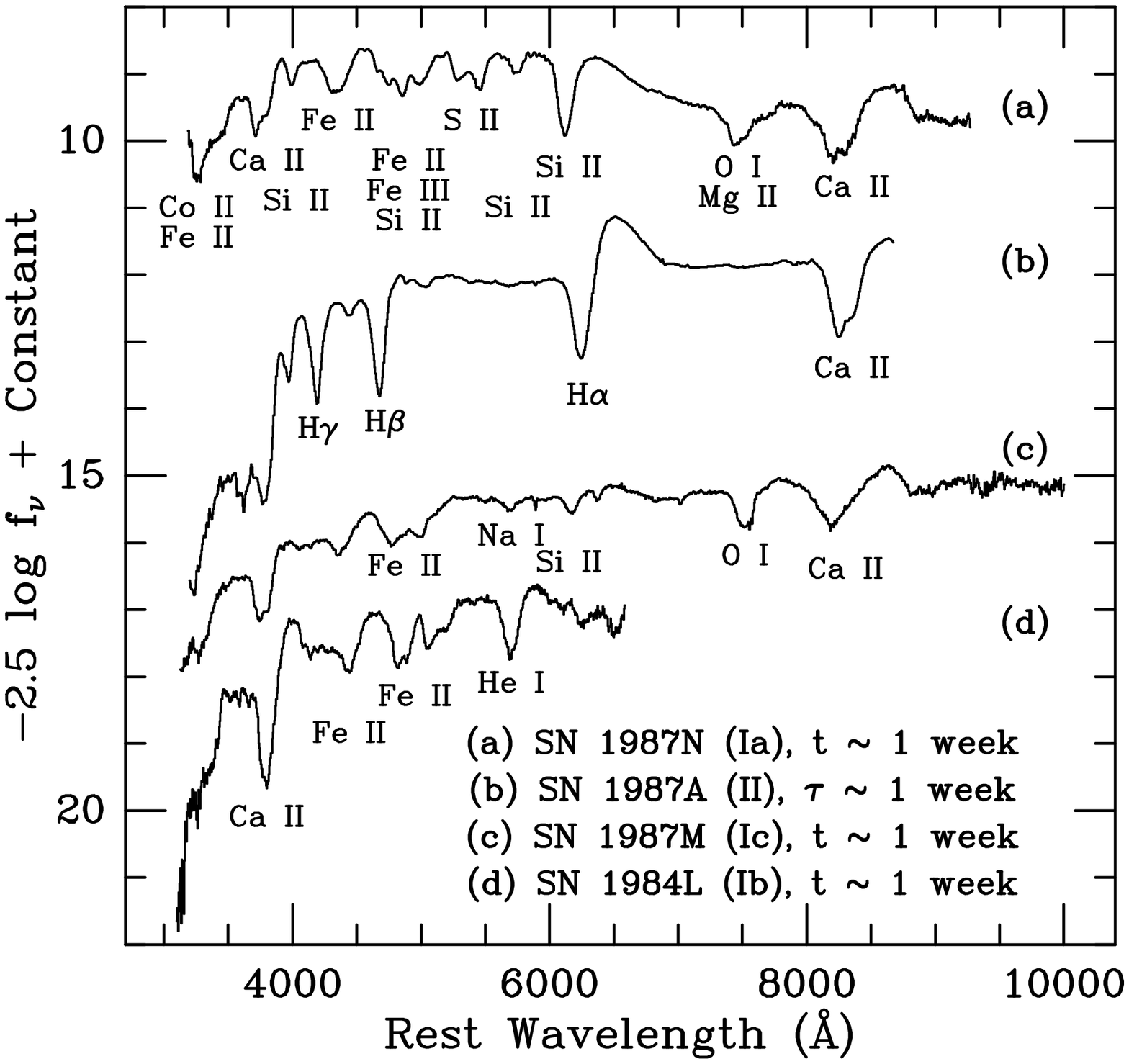,height=4.5truein,width=4.5truein,angle=0}
}
\noindent
{\it Figure 1:} Early-time spectra of SNe, showing the main subtypes.\\
\medskip

   The late-time ($t \gtrsim 4$ months) optical spectra of SNe provide
additional constraints on the classification scheme (Figure 2). SNe~Ia show
blends of dozens of Fe emission lines, mixed with some Co lines.  SNe~Ib and
Ic, on the other hand, have relatively unblended emission lines of
intermediate-mass elements such as O and Ca. At this phase, SNe~II are
dominated by the strong H$\alpha$ emission line; in other respects, most of
them spectroscopically resemble SNe~Ib and Ic, but with narrower emission
lines. The late-time spectra of SNe~II show substantial heterogeneity, as do
the early-time spectra.

  To a first approximation, the light curves of SNe~I are all broadly similar
\cite{lei91a}. SNe~Ib usually have slower decline rates than SNe~Ic; however,
SNe~Ic may come in ``slow" and ``fast" varieties \cite{clo97a,clo97b}.  The
light curves of SNe~II exhibit much dispersion \cite{pat93}, though it is
useful to subdivide the majority of them into two relatively distinct
subclasses \cite{bar79,dog85}.  The light curves of SNe~II-L (``linear'')
generally resemble those of SNe~I, with a steep decline after maximum
brightness followed by a slower exponential tail. In contrast, SNe~II-P
(``plateau") remain within $\sim 1$ mag of maximum brightness for an extended
period. The light curve of SN 1987A, albeit atypical, was generically related to
those of SNe~II-P.

\bigskip

\hbox{
\hskip +0.7truein
\psfig{figure=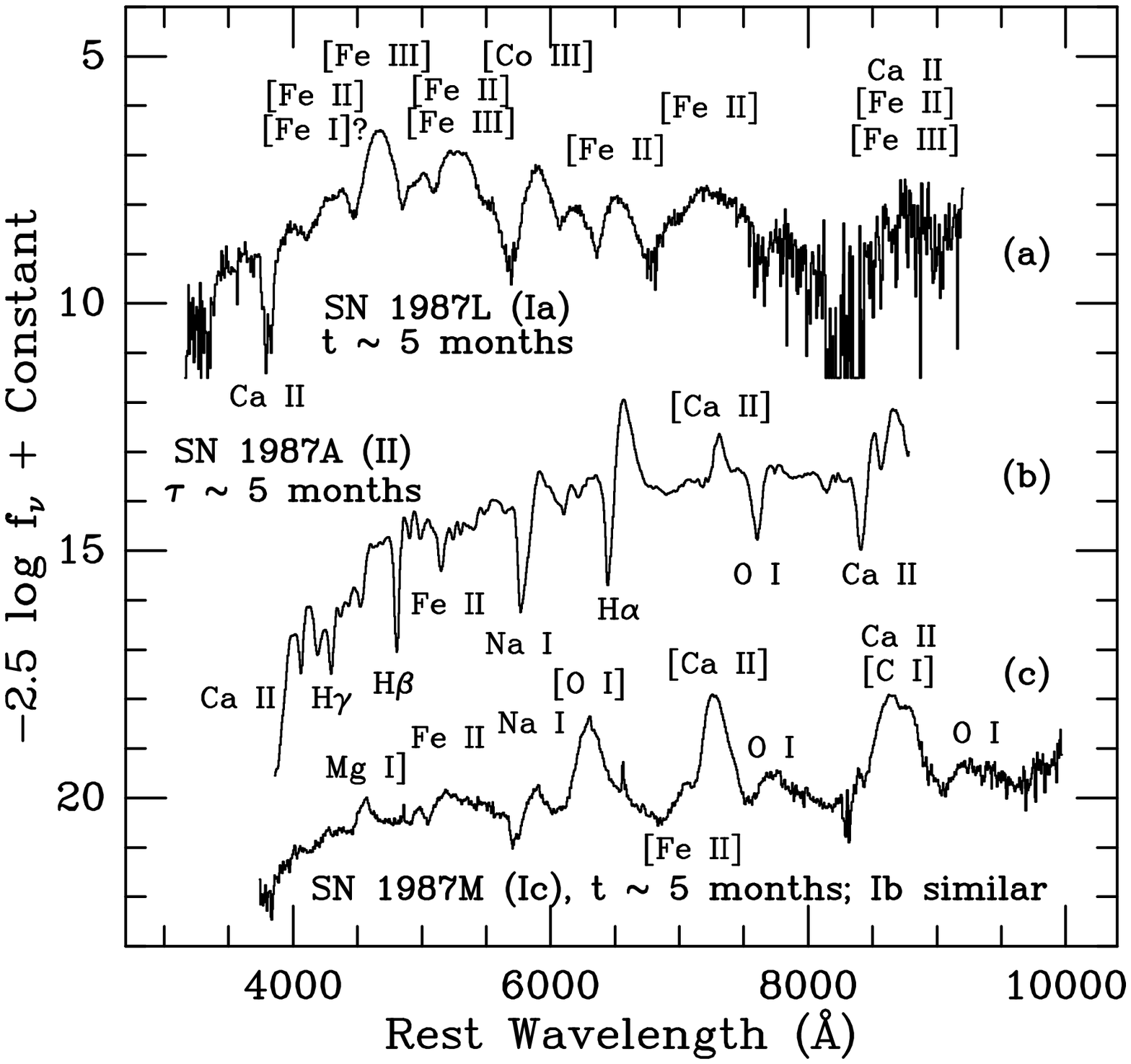,height=4.5truein,width=4.5truein,angle=0}
}

\noindent
{\it Figure 2:} Late-time spectra of SNe.  At even later phases, SN
1987A was dominated by strong emission lines of H$\alpha$, [O~I], [Ca~II], and
the Ca~II near-infrared triplet.
\medskip

   The locations at which SNe occur provide important clues to their nature,
and to the mass of their progenitor stars. SNe~II, Ib, and Ic have {\it never}
been seen in elliptical galaxies, and rarely if ever in S0 galaxies. They are
generally in or near spiral arms and H~II regions \cite{van96}, implying that
their progenitors must have started their lives as massive stars ($\gtrsim
10~M_\odot$). The progenitors of SNe~II are thought to suffer core collapse and
subsequently ``rebound" with help from neutrinos \cite{arn89,bur00}, leaving a
neutron star or perhaps in some cases a black hole \cite{bro94}. Most workers
now believe that SNe~Ib/Ic are produced by the same mechanism as SNe~II, except
that the progenitors were stripped of their hydrogen (SN~Ib) and possibly
helium (SN~Ic) envelopes prior to exploding, either via mass transfer to
companion stars \cite{nom94,woo95} or through winds (e.g., \cite{woo93,swa93}).
White dwarf models have been discussed \cite{bra86} but are unlikely.

\bigskip

\hbox{
\hskip +1truein
\psfig{figure=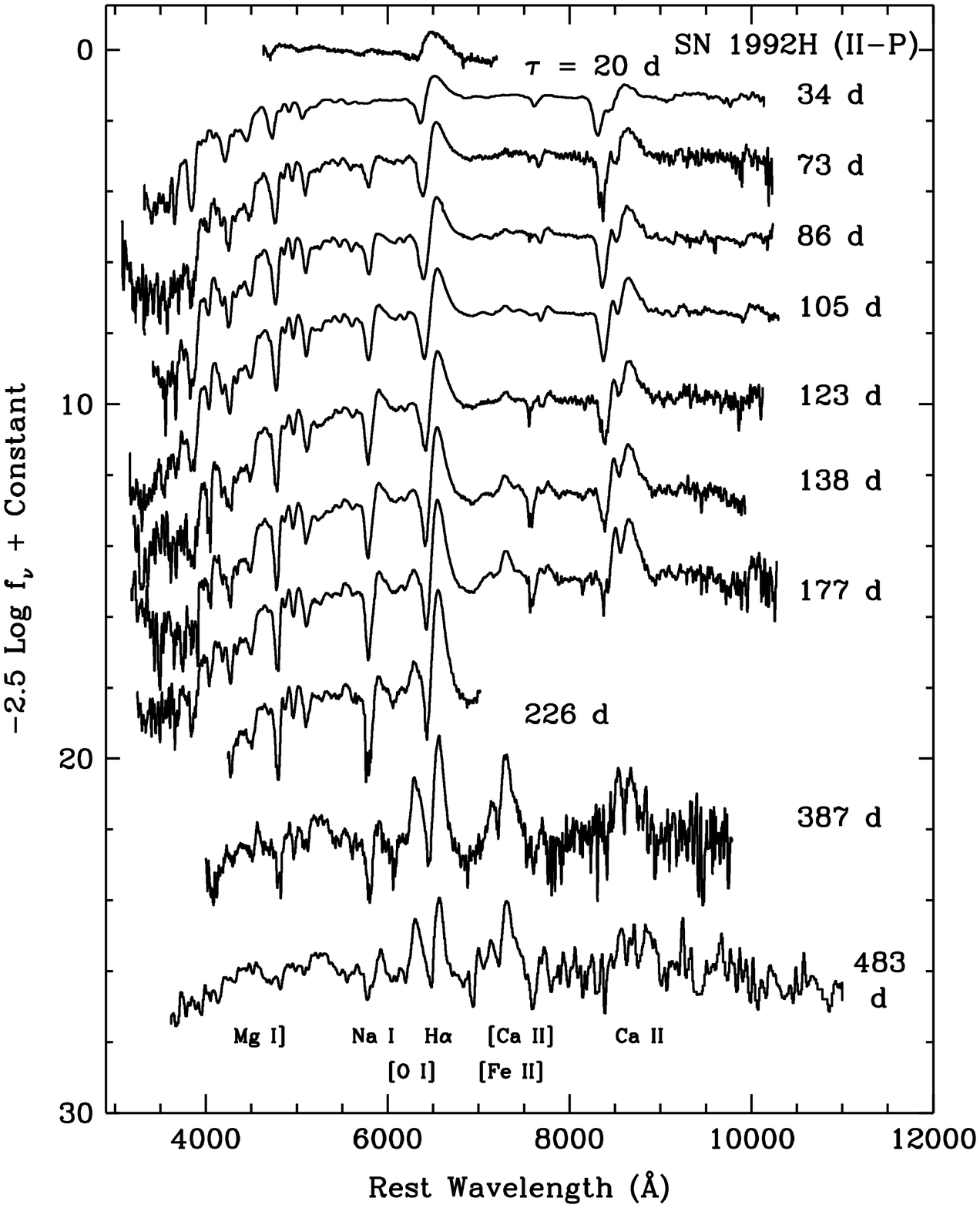,height=5truein,width=4truein,angle=0}
}

\noindent
{\it Figure 3:} Montage of spectra of SN 1992H in NGC 5377. Epochs
(days) are given relative to the estimated time of explosion,
February 8, 1992.
\medskip

\section*{ SUBCLASSES OF TYPE II SUPERNOVAE}

  Most SNe~II-P seem to have a relatively well-defined spectral development, as
shown in Figure 3 for SN 1992H (see also reference \cite{clo96}). At early
times the spectrum is nearly featureless and very blue, indicating a high color
temperature ($\gtrsim$ 10,000~K). He~I $\lambda$5876 with a P-Cygni profile
is sometimes visible. The temperature rapidly decreases with time, reaching
$\sim 5000$~K after a few weeks, as expected from the adiabatic expansion and
associated cooling of the ejecta. It remains roughly constant at this value
during the plateau (the photospheric phase), while the hydrogen recombination
wave moves through the massive ($\sim 10~M_\odot$) hydrogen ejecta and releases
the energy deposited by the shock. At this stage strong Balmer lines and Ca~II
H\&K with well-developed P-Cygni profiles appear, as do weaker lines of Fe~II,
Sc~II, and other iron-group elements. The spectrum gradually takes on a nebular
appearance as the light curve drops to the late-time tail; the continuum fades,
but H$\alpha$ becomes very strong, and prominent emission lines of [O~I],
[Ca~II], and Ca~II also appear.

   Few SNe~II-L have been observed in as much detail as SNe~II-P.  Figure 4
shows the spectral development of SN 1979C \cite{bra81}, an unusually luminous
member of this subclass. Near maximum brightness the spectrum is very blue and
almost featureless, with a slight hint of H$\alpha$ emission. A week later,
H$\alpha$ emission is more easily discernible, and low-contrast P-Cygni
profiles of Na~I, H$\beta$, and Fe~II have appeared. By $t \approx 1$ month,
the H$\alpha$ emission line is very strong but still devoid of an absorption
component, while the other features clearly have P-Cygni profiles. Strong,
broad H$\alpha$ emission dominates the spectrum at $t \approx 7$ months, and
[O~I] $\lambda\lambda$6300, 6364 emission is also present.  Several authors
\cite{whe90,avf91a,sch96} have speculated that the absence of H$\alpha$
absorption spectroscopically differentiates SNe~II-L from SNe~II-P, but the
small size of the sample of well-observed objects precluded definitive
conclusions.

\hbox{
\hskip +1truein
\psfig{figure=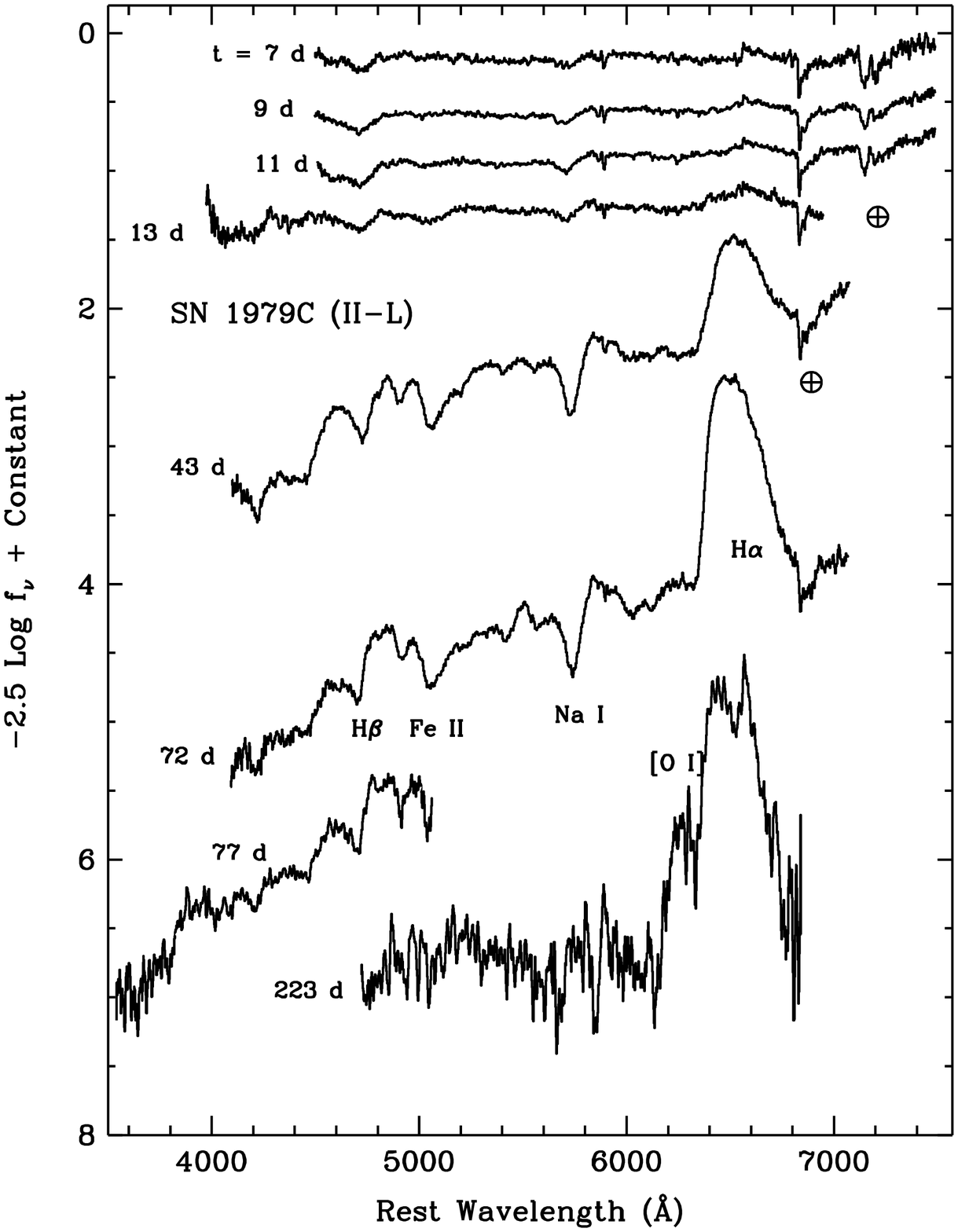,height=5truein,width=4truein,angle=0}
}

\noindent
{\it Figure 4:} Montage of spectra of SN 1979C in NGC 4321, from
reference \cite{bra81}; reproduced with permission. Epochs
(days) are given relative to the date of maximum brightness,
April 15, 1979.
\medskip

   The progenitors of SNe~II-L are generally believed to have relatively
low-mass hydrogen envelopes (a few $M_\odot$); otherwise, they would exhibit
distinct plateaus, as do SNe~II-P. On the other hand, they may have more
circumstellar gas than do SNe~II-P, and this could give rise to the
emission-line dominated spectra. They are often radio sources \cite{sra90};
moreover, the ultraviolet excess (at $\lambda \lesssim 1600$~\AA) seen in SNe
1979C and 1980K may be produced by inverse Compton scattering of photospheric
radiation by high-speed electrons in shock-heated ($T \approx 10^9$~K)
circumstellar material \cite{fra82,fra84}. Finally, the light curves of some
SNe~II-L reveal an extra source of energy: after declining exponentially for
several years, the H$\alpha$ flux of SN 1980K reached a steady level, showing
little if any decline thereafter \cite{uom86,lei91b}.  The excess almost
certainly comes from the kinetic energy of the ejecta being thermalized and
radiated due to an interaction with circumstellar matter \cite{che90,lei94}.

   The very late-time optical recovery of SNe 1979C and 1980K
\cite{lei91b,fes95,fes99} and other SNe~II-L supports the idea of ejecta
interacting with circumstellar material.  The spectra consist of a few strong,
broad emission lines such as H$\alpha$, [O~I] $\lambda\lambda$6300, 6364, and
[O~III] $\lambda\lambda$4959, 5007.  A {\it Hubble Space Telescope (HST)}
ultraviolet spectrum of SN 1979C reveals some prominent, double-peaked emission
lines with the blue peak substantially stronger than the red, suggesting dust
extinction within the expanding ejecta \cite{fes99}. The data show general
agreement with the emission lines expected from circumstellar interaction
\cite{che94}, but the specific models that are available show several
differences with the observations. For example, we find higher electron
densities ($10^5$ to $10^7$ cm$^{-3}$), resulting in stronger collisional
de-excitation than assumed in the models. These differences can be used to
further constrain the nature of the progenitor star. Note that based on
photometry of the stellar populations in the environment of SN 1979C (from {\it
HST} images), the progenitor of the SN was at most 10 million years years old,
so its initial mass was probably 17--18~$M_\odot$ \cite{van99}.

   During the past decade, there has been the gradual emergence of a new,
distinct subclass of SNe~II \cite{avf91a,avf91b,sch90,lei94} whose ejecta are
believed to be {\it strongly} interacting with dense circumstellar gas, even at
early times (unlike SNe~II-L). The derived mass-loss rates for the progenitors
can exceed $10^{-4} M_\odot$ yr$^{-1}$ \cite{chu94}.  In these objects, the
broad absorption components of all lines are weak or absent throughout their
evolution.  Instead, their spectra are dominated by strong emission lines, most
notably H$\alpha$, having a complex but relatively narrow profile. Although the
details differ among objects, H$\alpha$ typically exhibits a very narrow
component (FWHM $\lesssim 200$ km s$^{-1}$) superposed on a base of
intermediate width (FWHM $\approx$ 1000--2000 km s$^{-1}$; sometimes a very
broad component (FWHM $\approx$ 5000--10,000 km s$^{-1}$) is also present. This
subclass was christened ``Type IIn" \cite{sch90}, the ``n" denoting ``narrow"
to emphasize the presence of the intermediate-width or very narrow emission
components. Representative spectra of five SNe~IIn are shown in Figure 5, with
two epochs for SN 1994Y.

   The early-time continua of SNe~IIn tend to be bluer than normal.
Occasionally He~I emission lines are present in the first few spectra (e.g., SN
1994Y in Figure 5). Very narrow Balmer absorption lines are visible in the
early-time spectra of some of these objects, often with corresponding Fe~II,
Ca~II, O~I, or Na~I absorption as well (e.g., SNe 1994W and 1994ak in Figure
5). Some of them are unusually luminous at maximum brightness, and they
generally fade quite slowly, at least at early times. The equivalent width of
the intermediate H$\alpha$ component can grow to astoundingly high values at
late times. The great diversity in the observed characteristics of SNe~IIn
provides clues to the various degrees and forms of mass loss late in the lives
of massive stars.

\hbox{
\hskip +0.6truein
\psfig{figure=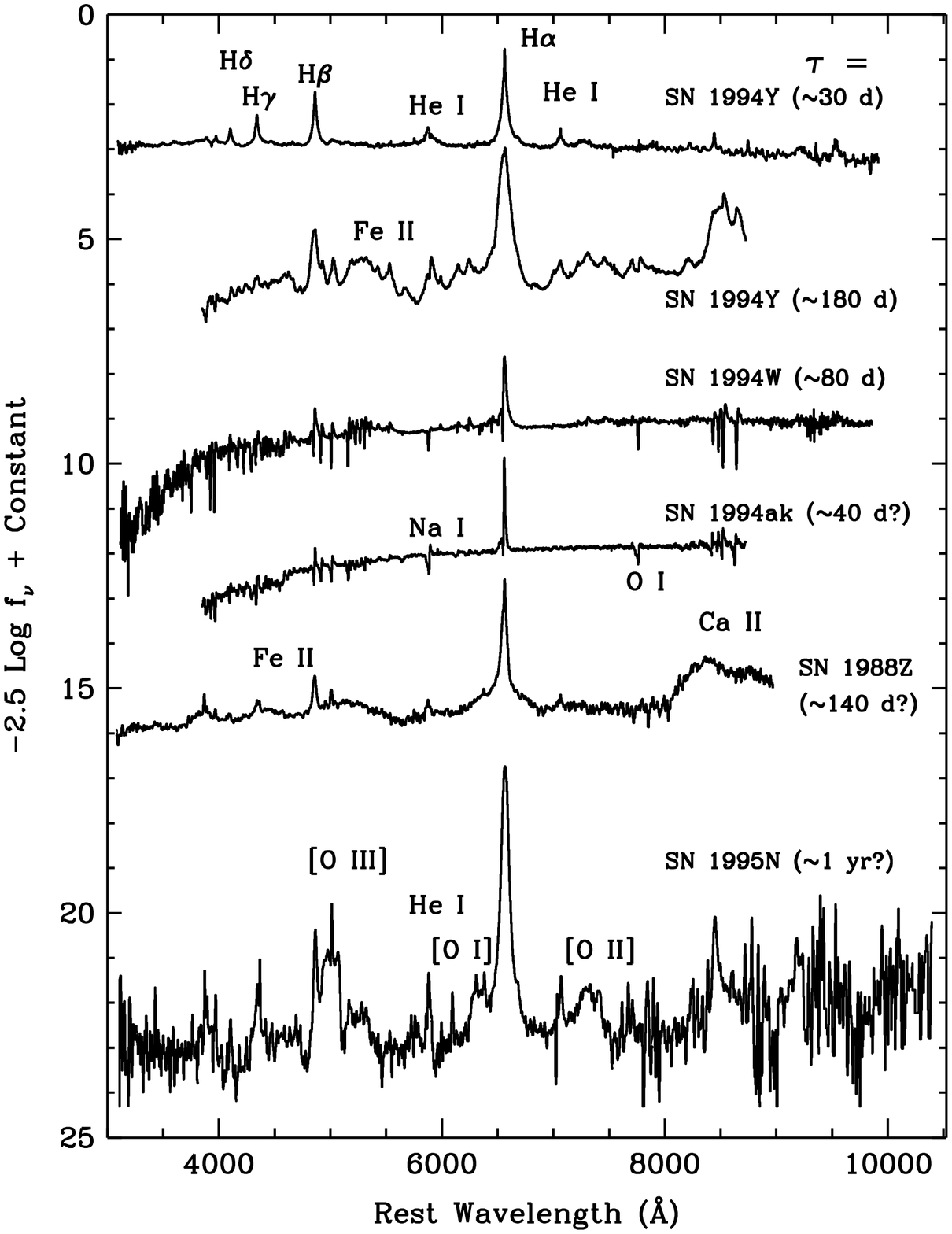,height=6.5truein,width=5truein,angle=0}
}
\bigskip
\noindent
{\it Figure 5:} Montage of spectra of SNe~IIn. Epochs are given relative 
to the estimated dates of explosion.\\

\section*{ TYPE II SUPERNOVA IMPOSTORS?}

  The peculiar SN~IIn 1961V (``Type V" according to Zwicky \cite{zwi65}) had
probably the most bizarre light curve ever recorded. (SN 1954J, also known as
``Variable 12" in NGC 2403, was similar \cite{hum94}.)  Its progenitor was a
very luminous star, visible in many photographs of the host galaxy (NGC 1058)
prior to the explosion. Perhaps SN 1961V was not a genuine supernova (defined
to be the violent destruction of a star at the end of its life), but rather the
super-outburst of a luminous blue variable such as $\eta$ Carinae
\cite{goo89,avf95}.

   A related object may have been SN~IIn 1997bs, the first SN discovered in the
Lick Observatory Supernova Search (LOSS, described later in this review).
Its spectrum was peculiar (Figure 6), consisting of narrow Balmer
and Fe~II emission lines superposed on a featureless continuum. Its progenitor
was discovered in an {\it HST} archival image of the host galaxy
\cite{van00}. It is a very luminous star ($M_V \approx -7.4$ mag), and it
didn't brighten as much as expected for a SN explosion ($M_V \approx -13$ at
maximum). These data suggest that SN 1997bs may have been like SN 1961V ---
that is, a supernova impostor. The real test will be whether the star is still
visible in future {\it HST} images obtained years after the outburst.

\hbox{
\hskip +0.3truein
\psfig{figure=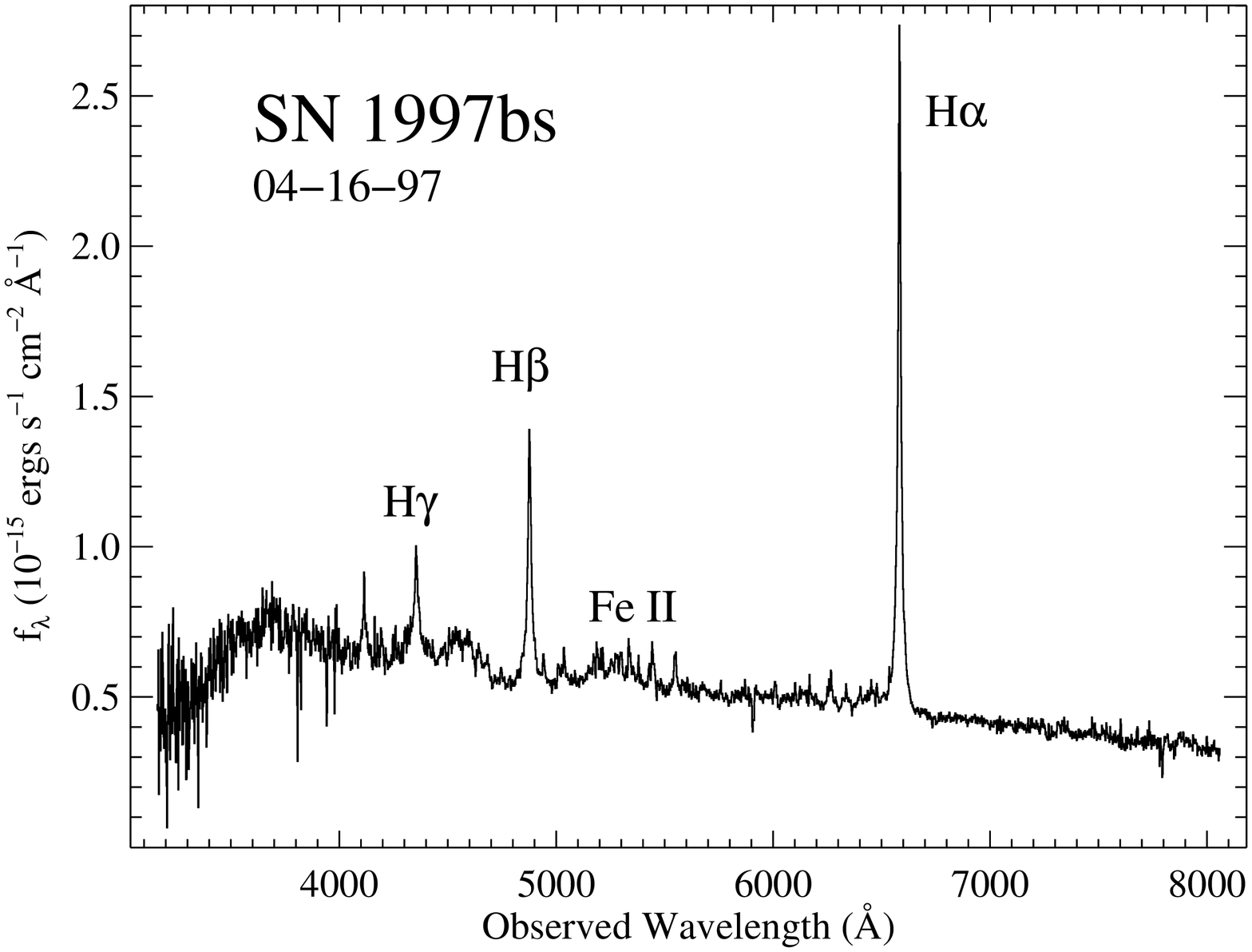,height=5truein,width=3truein,angle=90}
}
\noindent
{\it Figure 6:} Spectrum of SN 1997bs, obtained on April 16, 1997 UT.\\

\section*{ TYPE Ib AND Ic SUPERNOVAE}

  In the 1960s, Bertola and collaborators \cite{ber64,ber65} recognized that
not all SNe~I are of the ``classical" variety (now known as SNe~Ia) with a
strong absorption trough near 6150~\AA. By the mid-1980s these objects came to
be known as SNe~Ib \cite{eli85}, and He~I lines were identified in their
early-time spectra \cite{har87}.  Gradually it became clear that SNe~Ib
constitute a heterogeneous subclass, with substantial variations in the
observed He~I strengths in spectra obtained around maximum brightness. Wheeler
\& Harkness (\cite{whe86}; see also \cite{har87}) suggested that SNe~Ib should
actually be divided into two separate categories: SNe~Ib are those showing
strong He~I absorption lines (especially He~I $\lambda$5876) in their
early-time photospheric spectra, while SNe~Ic are those in which He~I is not
easily discernible. However, they modeled SNe~Ic in the same physical way as
SNe~Ib \cite{whe87}, but with different relative concentrations of He
and O in the envelope.

   A large, comprehensive study of SNe~Ib and SNe~Ic was recently completed by
my group \cite{mat01}. The relative depths of the helium absorption lines in
the spectra of the SNe~Ib appear to provide a measurement of the temporal
evolution of the supernova, with He~I $\lambda$5876 and He~I $\lambda$7065
growing in strength relative to He~I $\lambda$6678 over time.  Some SNe~Ic show
evidence for weak He~I absorption, but most do not.  Aside from the presence or
absence of the helium lines, there are other spectroscopic differences between
SNe~Ib and SNe~Ic.  On average, the O~I $\lambda$7774 line is stronger in
SNe~Ic than in SNe~Ib.  In addition, the SNe~Ic have distinctly broader
emission lines at late times, indicating either a consistently larger explosion
energy and/or a lower envelope mass for SNe~Ic than for SNe~Ib. These results
are consistent with the idea that the progenitors of SNe~Ic are massive stars
that have lost more of their envelope (i.e., much of the helium layer) than the 
progenitors of SNe~Ib. The general hypothesis that SNe~Ib/Ic have ``stripped"
progenitors is greatly supported by the discovery of links between SNe~II and
SNe~Ib/Ic, as I discuss next.

\section*{ LINKS BETWEEN TYPE II AND TYPE Ib/Ic SUPERNOVAE}

   Filippenko \cite{avf88} presented the case of SN 1987K, which appeared to be
a link between SNe~II and SNe~Ib. Near maximum brightness, it was undoubtedly a
SN~II, but with rather weak photospheric Balmer and Ca~II lines. Many months
after maximum brightness, its spectrum was essentially that of a SN~Ib. The
simplest interpretation is that SN 1987K had a meager hydrogen atmosphere at
the time it exploded; it would naturally masquerade as a SN~II for a while, and
as the expanding ejecta thinned out the spectrum would become dominated by
emission from deeper and denser layers. The progenitor was probably a star
that, prior to exploding via iron core collapse, lost almost all of its
hydrogen envelope either through mass transfer onto a companion or as a result
of stellar winds.  Such SNe were dubbed ``SNe~IIb" by Woosley et
al. \cite{woo87}, who had proposed a similar preliminary model for SN 1987A
before it was known to have a massive hydrogen envelope.

   The data for SN 1987K (especially its light curve) were rather sparse,
making it difficult to model in detail.  Fortunately, the Type II SN 1993Jin
NGC 3031 (M81) came to the rescue, and was studied in greater detail than any
supernova since SN 1987A \cite{whe96}.  Its light curves \cite{ric96} and
spectra \cite{avf93,avf94,mat00a,mat00b} amply supported the hypothesis that
the progenitor of SN 1993J probably had a low-mass (0.1--0.6~$M_\odot$)
hydrogen envelope above a $\sim 4~M_\odot$ He core
\cite{nom93,pod93,woo94}. Figure 7 shows several early-time spectra of SN
1993J, showing the emergence of He~I features typical of SNe~Ib. Considerably
later (Figure 8), the H$\alpha$ emission nearly disappeared, and the spectral
resemblance to SNe~Ib was strong.  The general consensus is that its initial
mass was $\sim 15~M_\odot$. A star of such low mass cannot shed nearly its
entire hydrogen envelope without the assistance of a companion star. Thus, the
progenitor of SN 1993J probably lost most of its hydrogen through mass transfer
to a bound companion 3--20~AU away. In addition, part of the gas may have been
lost from the system.  Had the progenitor lost essentially {\it all} of its
hydrogen prior to exploding, it would have had the optical characteristics of
SNe~Ib. There is now little doubt that most SNe~Ib, and probably SNe~Ic as
well, result from core collapse in stripped, massive stars, rather than from
the thermonuclear runaway of white dwarfs.

\hbox{
\hskip +0.3truein
\psfig{figure=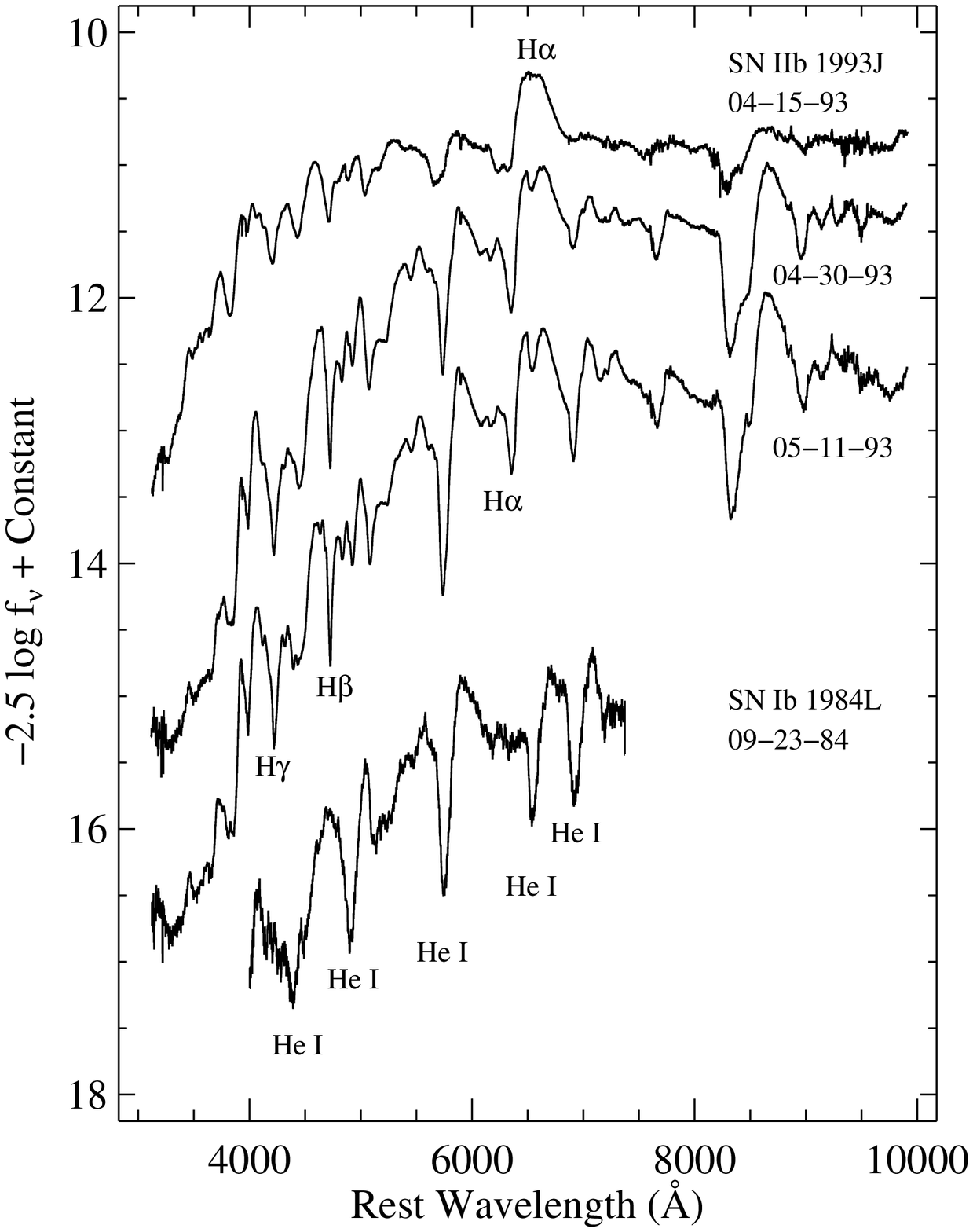,height=4.6truein,width=5truein,angle=0}
}
\bigskip
\noindent
{\it Figure 7:} Early-time spectral evolution of SN 1993J. A comparison
with the Type Ib SN 1984L is shown at bottom, demonstrating the
presence of He~I lines in SN 1993J. The explosion date was
March 27.5, 1993.\\

   SN 1993J held several more surprises. Observations at radio \cite{van94} and
X-ray \cite{suz95} wavelengths revealed that the ejecta are interacting with
relatively dense circumstellar material \cite{fra96}, probably ejected from the
system during the course of its pre-SN evolution. Optical evidence for this
interaction also began emerging at $\tau \gtrsim 10$ months: the H$\alpha$
emission line grew in relative prominence, and by $\tau \approx 14$ months it
had become the dominant line in the spectrum \cite{avf94,pat95,fin95},
consistent with models \cite{che94}.  Its profile was very broad (FWHM
$\approx$ 17,000 km s$^{-1}$; Figure 8) and had a relatively flat top, but with
prominent peaks and valleys whose likely origin is Rayleigh-Taylor
instabilities in the cool, dense shell of gas behind the reverse shock
\cite{che92}. Radio VLBI measurements show that the ejecta are circularly
symmetric, but with significant emission asymmetries \cite{mar95}, possibly
consistent with the asymmetric H$\alpha$ profile seen in some of the spectra
\cite{avf94}.

\hbox{
\hskip +0.3truein
\psfig{figure=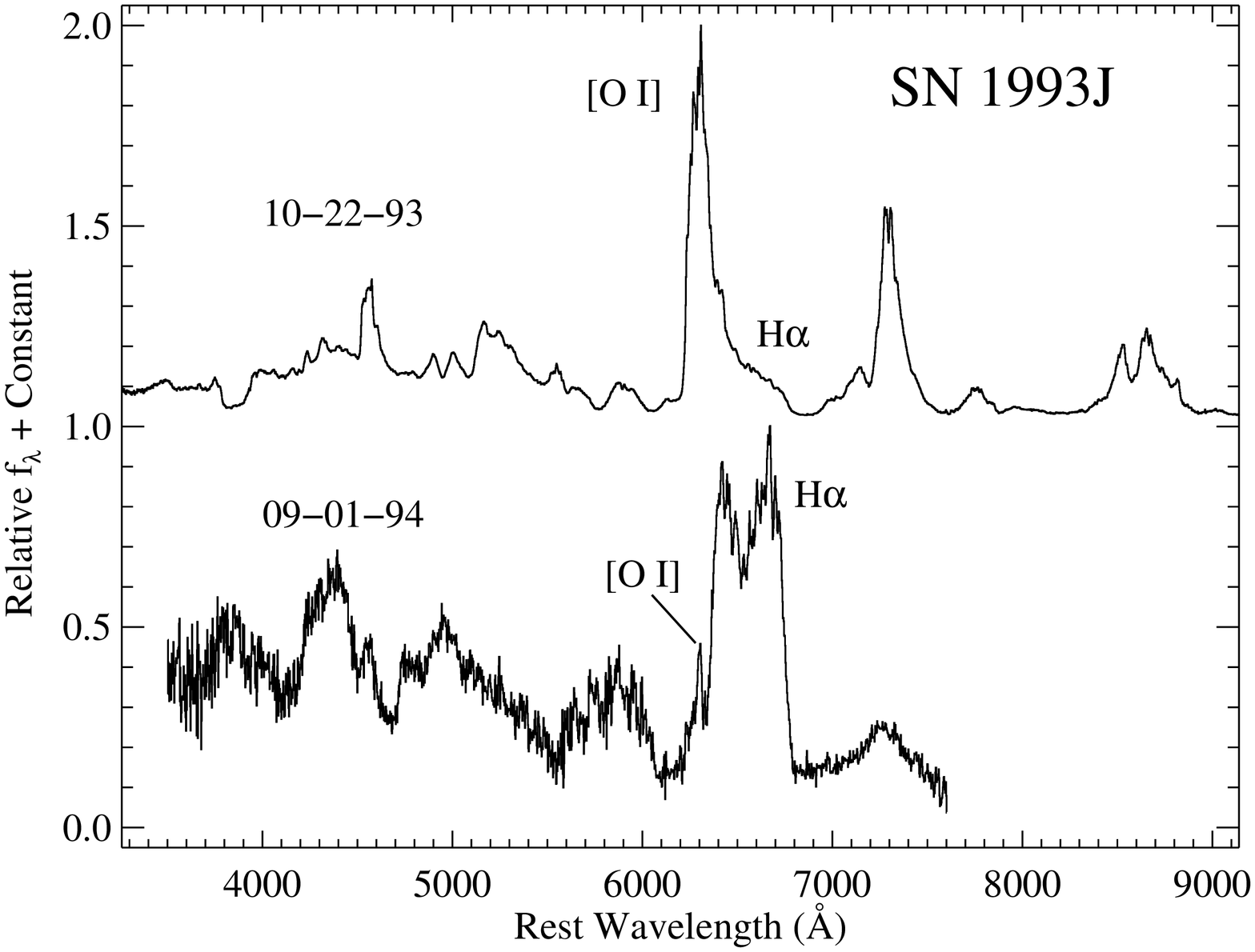,height=5truein,width=3truein,angle=90}
}
\bigskip
\noindent
{\it Figure 8:} In the {\it top} spectrum, which shows SN 1993J about 7
months after the explosion, H$\alpha$ emission is very weak; the
resemblance to spectra of SNe~Ib is striking. A year later ({\it bottom}), 
however, H$\alpha$ was once again the dominant feature
in the spectrum (which was scaled for display purposes).\\

\section*{ SPECTROPOLARIMETRY OF SUPERNOVAE}

   Spectropolarimetry of SNe can be used to probe their geometry.
The basic question is whether SNe are round, and the idea is
simple: A hot young SN atmosphere is dominated by electron scattering, which by
its nature is highly polarizing.  For an unresolved, spherical source, however,
the directional components of the electric vectors cancel exactly, yielding
zero net linear polarization.  Conversely, if the source is aspherical,
incomplete cancellation occurs and a net linear polarization results, with
typical continuum polarizations of $\sim 1\%$ expected for objects with
moderate ($\sim 20\%$) asphericity (e.g., \cite{hof91}), although the exact
amount of polarization observed is also sensitive to the viewing angle.
Spectropolarimetry is important for a full understanding of the physics of SN
explosions and can also provide information on the circumstellar environment of
SNe.

   My group obtained spectropolarimetry of one object from each of the major SN
types and subtypes \cite{leo00a}, generally with the Keck-II 10-m telescope
(but in a few important cases with the Lick 3-m Shane reflector). Most of the
objects exhibit a change in both the magnitude and direction of the
polarization across strong lines, especially the absorption troughs of the
strongest P-Cygni lines. This may result from global asymmetry of the
electron-scattering atmosphere and/or the underlying continuum region
\cite{mcc84,hof91}.
  
   We have studied SN~IIn 1998S in some detail \cite{leo00b} (see also
\cite{wan01}); its optical spectrum is dominated by strong, multi-component
emission lines, thought to be produced by an intense interaction between the
supernova and its dense circumstellar environment.  We combined our early-time
(3 days after discovery) spectropolarimetric observation with total flux
spectra spanning nearly 500 days. We measure an intrinsic continuum
polarization of $p \approx 3\%$ (one of the highest yet found for a SN),
suggesting a global asphericity of $\gtrsim 45\%$ from the models of
H\"{o}flich \cite{hof91}. The line profiles favor a ring-like geometry for the
circumstellar gas, generically similar to what is seen directly in SN 1987A,
although much denser and closer to the progenitor in SN 1998S.

   We also found that one month after exploding, the Type IIb SN 1993J had a
polarized flux spectrum resembling spectra of SNe~Ib, with prominent He~I lines
\cite{tra97}.  The data are consistent with models in which the polarization is
produced by an asymmetric He core configuration of material.  It is interesting
that the percent polarization may increase with decreasing envelope mass, along
the sequence Type II, IIb, Ib, and Ic \cite{whe00,wan01,leo01b}, suggesting that
asymmetries in massive stars become more pronounced as one probes deeper into
the core.
 
   SN 1999em, an extremely bright ($m_V \approx 13.5$) SN II-P, provided the
rare opportunity to study the geometry of a ``normal'' core-collapse event at
multiple epochs \cite{leo01a}.  We obtained spectropolarimetry at 3
plateau-phase epochs, and then one final epoch long after it had dropped off
the plateau (Fig. 9).  A very low but temporally increasing polarization level
suggests a substantially spherical geometry at early times that becomes more
aspherical at late times as ever-deeper layers of the ejecta are revealed.  We
speculate that the thick hydrogen envelope intact at the time of explosion in
SNe II-P might serve to dampen the effects of an intrinsically aspherical
explosion. The increase in asphericity seen at later times is consistent with
the trend identified above among stripped-envelope core-collapse SNe: the
deeper we peer, the more evidence we find for asphericity.  The natural
conclusion that it is {\it explosion} asymmetry that is responsible for the
polarization has fueled the idea that some core-collapse SNe produce gamma-ray
bursts (GRB; e.g., \cite{blo99}) through the action of a jet of material
aimed fortuitously at the observer, the result of a ``bipolar,'' jet-induced,
SN explosion \cite{kho99,whe00}.

\bigskip

\hbox{
\hskip +1truein
\psfig{figure=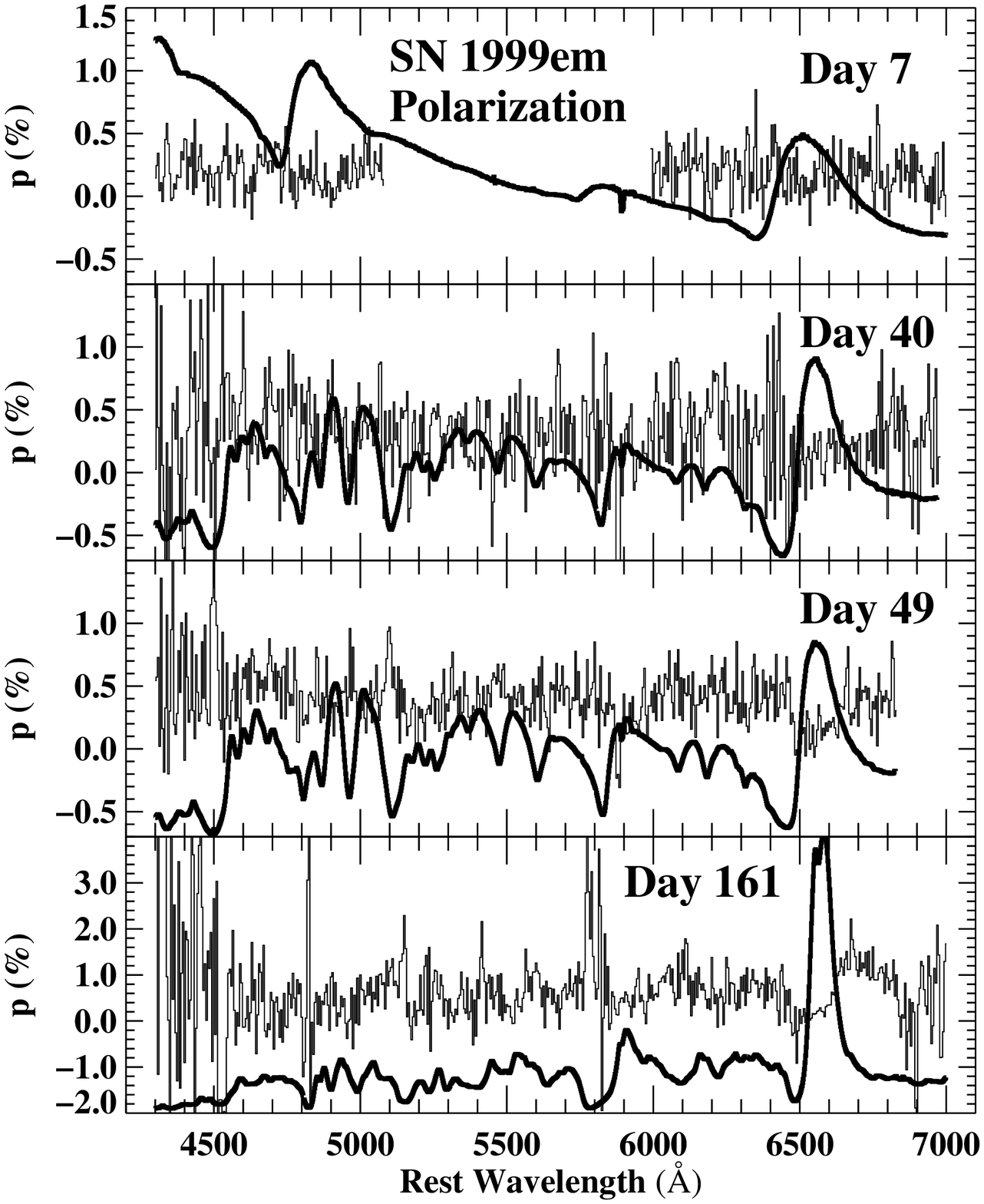,height=5truein,width=4truein,angle=0}
}
\bigskip
\noindent
{\it Figure 9:} Montage of the observed optical polarization of SN 1999em, with
arbitrarily scaled total flux spectra {\it (thick lines)} overplotted for
comparison of features. The polarization spectra are binned 5~\AA\ bin$^{-1}$
to improve the S/N ratio.  See reference \cite{leo01b} for details.\\

\section*{ SUPERNOVAE ASSOCIATED WITH GAMMA-RAY BURSTS?}

  At least a small fraction of gamma-ray bursts (GRBs) may be associated with
nearby SNe. Probably the most compelling example thus far is that of SN 1998bw
and GRB 980425 (e.g., \cite{gal98,iwa98,woo99,sta00}), which were temporally
and spatially coincident. SN 1998bw was, in many ways, an extraordinary SN; it
was very luminous at optical and radio wavelengths, and it showed evidence for
relativistic outflow. Its bizarre optical spectrum is often classified as that
of a SN~Ic, but the object should be called a ``peculiar SN~Ic" if not a
subclass of its own; the spectrum was distinctly different from that of a
normal SN~Ic.

   Models suggest that SNe associated with GRBs are highly asymmetric; thus,
spectropolarimetry should provide some useful tests. In particular, perhaps
objects such as SN 1998S, mentioned above, would have been seen as GRBs had
their rotation axis been pointed in our direction. That of SN 1998S was almost
certainly {\it not} aligned with us \cite{leo00b}; both the spectropolarimetry
and the appearance of double-peaked H$\alpha$ emission suggest an inclined
view, rather than pole-on.

\bigskip

\hbox{
\hskip +0.3truein
\psfig{figure=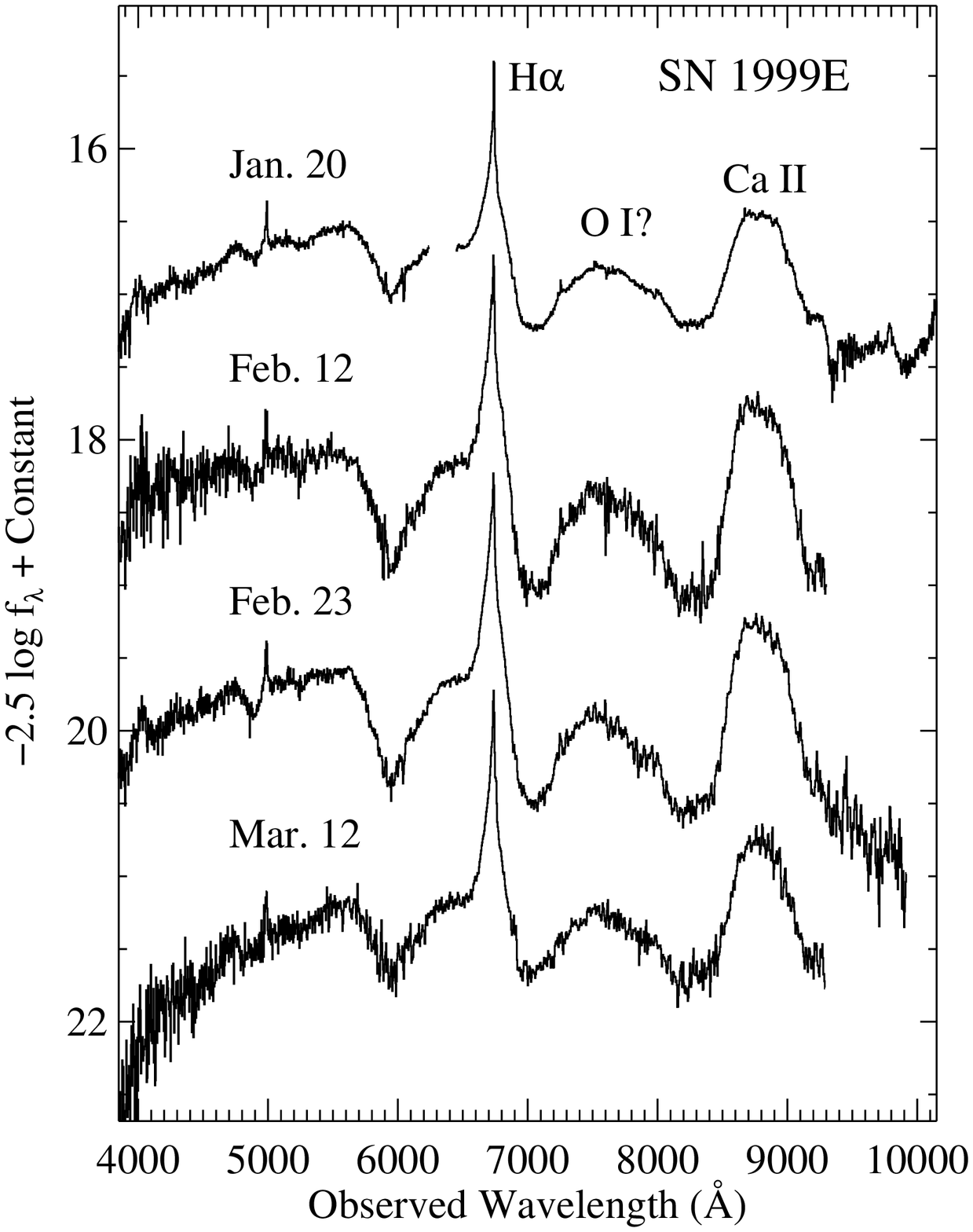,height=5truein,width=4.5truein,angle=0}
\hskip +.3truein
}
\noindent
{\it Figure 10:} Spectral evolution of SN 1999E, which may have been
associated with GRB 980910.
\medskip

   The case of GRB 970514 and the very luminous SN IIn 1997cy is also
interesting \cite{ger00,tur00}; there is a reasonable possibility that the two
objects were associated. The optical spectrum of SN 1997cy was highly unusual,
and bore some resemblance to that of SN 1998bw, though there were some
differences as well. SN 1999E, which might be linked with GRB 980910 but with
large uncertainties \cite{tho99}, also had an optical spectrum similar to that
of SN 1997cy \cite{avf99,cap99a}; see Figure 10.  The undulations are very
broad, indicating high ejection velocities. Besides H$\alpha$, secure line
identifications are difficult, though some of the emission features seem to be
associated with oxygen and calcium. Perhaps SN 1999E was produced by the highly
asymmetric collapse of a carbon-oxygen core.

\section*{ SUPERNOVA RATES}

   I conclude with a brief summary of supernova rates. The derived rates of
various types of SNe as a function of Hubble type are still quite uncertain; no
single search has found a sufficiently large sample of SNe, and combining
different searches can be misleading due to differences in the dominant
selection effects. Probably the best existing determination of supernova rates
is that of Capellaro et al. \cite{cap99b}, who combined the results of five searches
while being attentive to various selection effects. Four of the searches had
been done with photographic plates, while the fifth was a visual search by the
Rev. Robert Evans. Given that photographic searches can be rather insensitive
to SNe near the central regions of galaxies, and that the Evans visual search
did not go very deep, we must avoid the temptation to take the overall results
(see table below) too literally. Nevertheless, they can be used as a rough
guideline.

\begin{center}
SNu [\#(100 yr)$^{-1} (10^{10}L_\odot^B)^{-1}$] 
\end{center}
\begin{center}
\begin{tabular}{lcccc}
\hline\hline
Galaxy type & Ia & Ib/c & II & All \\
\hline
E-S0 & $0.18\pm0.06$ & $<0.01$ & $<0.02$ & $0.18\pm0.06$ \\
S0a-Sb & $0.18\pm0.07$ & $0.11\pm0.06$& $0.42\pm0.19$ & $0.72\pm0.21$\\
Sbc-Sd & $0.21\pm0.08$ & $0.14\pm0.07$ & $0.86\pm0.35$ & $1.21\pm0.37$ \\
Others$^*$ & $0.40\pm0.16$ & $0.22\pm0.16$ & $0.65\pm0.39$&$1.26\pm0.45$ \\
All & $0.20\pm0.06$ & $0.08\pm0.04$ & $0.40\pm0.19$ & $0.68\pm0.20$ \\
\hline
\end{tabular}
\end{center}

\hskip 0.5 truein \emph{Note:} $H_0 = 75$ km s$^{-1}$ Mpc$^{-1}$; scale by ($H_0/75)^2$.

\hskip 0.5 truein $^*$Others include Sm, Irr, Pec.

\bigskip

  My group is trying to remedy the situation by conducting a long-term search
for nearby SNe (with redshifts generally less than 5000 km s$^{-1}$) in a uniform
manner \cite{wli00,avf01}. Special emphasis is placed on finding them well
before maximum brightness. We are using the Katzman Automatic Imaging Telescope
(KAIT) at Lick Observatory, a fully robotic 0.75-m reflector equipped with a
CCD imaging camera.  Its telescope control system checks the weather, opens the
dome, points to the desired objects, finds and acquires guide stars (only for
long integrations), exposes, stores the data, and manipulates the data without
human intervention. There is a filter wheel with 20 slots, including
UBVRI. Five-minute guided exposures yield $R \approx 20$ mag.

  A limit of about 19 mag ($4\sigma$) is reached in the 25-second unfiltered,
unguided exposures of our Lick Observatory Supernova Search (LOSS). We observe
up to 1200 galaxies per night, and try to cycle back to the same galaxies after
3--4 nights.  Our software automatically subtracts new images from old ones and
identifies supernova candidates that are subsequently examined by undergraduate
research assistants. LOSS found 20 SNe in 1998, 40 in 1999, and 36 in 2000,
making KAIT the world's most successful search engine for nearby SNe. (Note
that we found SN 2000A and SN 2001A --- and hence the first supernova of the
new millennium regardless of one's definition of the turn of the millennium!)
After a few more years, we hope to have found enough SNe to provide a
meaningful update on the supernova rates given above.

  Multi-filter follow-up photometry is conducted of the most important SNe, and
all objects are monitored in unfiltered mode.  A Web page describing LOSS is at
http://astro.berkeley.edu/$\sim$bait/kait.html .  As a byproduct of LOSS, we
also find novae in the Local Group, comets, asteroids, and cataclysmic
variables.

\section*{ ACKNOWLEDGMENTS}

   My recent research on SNe has been financed by the NSF, most recently
through grant AST-9987438, as well as by NASA grants GO-7821, GO-8243, GO-8648,
AR-6371, and AR-8006 from the Space Telescope Science Institute, which is
operated by AURA, Inc., under NASA Contract NAS5-26555. Additional funding was
provided by NASA/Chandra grant GO-0-1009C. KAIT and its associated science have
been made possible with funding or donations from NSF, NASA, the Sylvia and Jim
Katzman Foundation, Sun Microsystems Inc., Lick Observatory, the
Hewlett-Packard Company, Photometrics Ltd., AutoScope Corporation, and the
University of California. This paper was written while I was supported by
a Guggenheim Fellowship.  I am grateful to the students and postdocs who have
worked with me on SNe over the past 15 years for their assistance and
discussions.

\end{document}